\documentclass{conm-p-l}
\usepackage{amsfonts}

\newtheorem{theorem}{Theorem}[section]
\newtheorem{lemma}[theorem]{Lemma}
\theoremstyle{definition}
\newtheorem{definition}[theorem]{Definition}

\theoremstyle{remark}
\newtheorem{remark}[theorem]{Remark}
\numberwithin{equation}{section}
\theoremstyle{plain}

\newtheorem{claim}{Claim}

\newtheorem{corollary}{Corollary}

\newtheorem{proposition}{Proposition}

\copyrightinfo{2001}{C. Korff}

\begin{document}
\title[Solving Baxter's TQ-equation]{Solving Baxter's TQ-equation via
representation theory}
\author{Christian Korff}
\address{School of Mathematics, University of Edinburgh, King's Buildings\\
Mayfield Road, Edinburgh EH9 3JZ, Scotland, UK}
\email{c.korff@ed.ac.uk}
\urladdr{http://www.maths.ed.ac.uk/\symbol{126}ckorff}
\thanks{}
\date{October, 2004}

\begin{abstract}
Baxter's TQ-equation is solved for the six-vertex model using the
representation theory of quantum groups at roots of unity. A novel
simplified construction of the Q-operator is given depending on a new free
parameter. Specializing this general construction to even roots of unity and
lattices with an odd number of columns two linearly independent operator
solutions of Baxter's TQ equation are obtained as special limits.
\end{abstract}

\maketitle

\section{Introduction}

Integrable lattice models of statistical mechanics form an important
interface between physics and mathematics and have given rise to numerous
discoveries relating both subjects, quantum groups are a celebrated example 
\cite{Drin, Jimbo}. Recent discoveries of infinite-dimensional non-abelian
symmetries at roots of unity in the case of the six-vertex model \cite{DFM}
and new developments in the eight-vertex model \cite{FM8v,FM8v2,FM8v3} have
opened up a number of questions of mathematical and physical interest. The
degeneracies in the spectrum of the transfer matrix at these particular
points and their relation with Bethe's ansatz \cite{Bethe} and Baxter's
TQ-equation \cite{Bx72,Bx73,BxBook} have been subject of a series of papers 
\cite{FM01a,FM01b,Bx02,Bx04}. Moreover, the case of primitive roots of unity
of order three has found particular interest because of its connection with
combinatorial aspects such as the enumeration of alternating-sign matrices
or plane partitions, see e.g. \cite{Ku96,RS,BGN}. Here Baxter's TQ-equation
has been explicitly solved for the groundstate eigenvalue of the six-vertex 
\cite{FSZ,S01} and more recently also for the eight-vertex model \cite%
{FM8v3,BM}.

In an independent development the techniques of the quantum inverse
scattering method \cite{QISM} have been applied to conformal field theory
and Baxter's TQ equation has been discussed in the context of the Liouville
model \cite{BLZ99, FKV}. This led to a representation theoretic approach of
constructing Baxter's $Q$-operator and in subsequent papers the method has
been generalized and applied to the six-vertex model away from a root of
unity \cite{AF97,RW02,KQ3}.

The representation theoretic construction of $Q$-operators for the
six-vertex model at roots of unity and their relation to the aforementioned
symmetries has been the main focus of the papers \cite{KQ,KQ2,KQ4}. 
\footnote{%
See also \cite{BS90} for a different construction and the related comments
in the introduction of \cite{KQ}.} It was demonstrated therein that the $Q$%
-operators constructed in \cite{KQ} provide an efficient tool to analyze the
degeneracies in the spectrum of the transfer matrix, reveal representation
theoretic information on the affine symmetry algebra and allow to determine
whether a second linear independent solution to Baxter's $TQ$ equation
exists. It is the last aspect which we will further highlight in this
article.

The possible existence of two linearly independent solutions to the
six-vertex model has been discussed in \cite{PrSt} excluding the root of
unity case. See also \cite{KWLZ} for the discussion of the eight-vertex
case. While the aforementioned papers addressed the question on the level of
eigenvalues, it has been shown in \cite{KQ4} that these solutions arise in
the spectrum of the $Q$-operators explicitly constructed in \cite{KQ} at
primitive roots of unity. The analogous discussion for the eight-vertex
model based on Baxter's $Q$-operator constructed in his 1972 work \cite{Bx72}
and numerical computations has been subject of the paper \cite{FM8v3}. There
is, however, a significant difference as Baxter's construction procedure for
the $Q$-operator does not rely on representation theory or contains free
parameters. It is the representation theory of quantum groups at roots of
unity used in \cite{KQ} which provides the key to the derivation of the $TQ$
equation and a second equally important functional equation, see equation (%
\ref{func}) in the text.

In comparison with the discussion contained in \cite{KQ4} the new aspect
presented here is the explicit construction of $Q$-operators which contain
one more free parameter. This puts one into the position to take two special
limits which yield the two linearly independent \emph{operator}-solutions of
Baxter's $TQ$ equation; see Definition 3.2 and Definition 3.5 in the text.
Based on these results we will show for even roots of unity and an odd
number of lattice columns that the proof of existence of two linear
independent operator-solutions to the $TQ$ equation and the completeness of
the Bethe ansatz above and below the equator can be reduced to a
well-defined mathematical problem: the construction of an intertwiner for
two quantum group representations at roots of unity. In the appendix we then
explicitly construct this intertwiner for the cases where the order of the
root of unity is four and six.

\section{Preliminaries}

Our definition of the six-vertex model and the discussion of Baxter's $TQ$%
-equation will be based on the Drinfel'd-Jimbo quantum group associated with
the affine algebra $\widehat{sl}_{2}$. We briefly recall some of its
defining relations in terms of the Chevalley-Serre basis; see e.g. \cite%
{CPbook} for further details.

\begin{proposition}
There is a quasi-triangular Hopf algebra $U_{q}(\widehat{sl}_{2})$ which is
generated by elements \{$e_{i},f_{i},q^{\pm h_{i}}$\}$_{i=0,1}$ obeying the
relations 
\begin{eqnarray}
q^{h_{i}}q^{h_{j}} &=&q^{h_{j}}q^{h_{i}},\quad
q^{h_{i}}q^{-h_{i}}=q^{-h_{i}}q^{h_{i}}=1,  \label{AQG} \\
q^{h_{i}}e_{j}q^{-h_{i}} &=&q^{\mathcal{A}_{ij}}e_{j},\quad
q^{h_{i}}f_{j}q^{-h_{i}}=q^{-\mathcal{A}_{ij}}f_{j}, \\
\lbrack e_{i},f_{j}] &=&\delta _{ij}~\frac{q^{h_{i}}-q^{-h_{i}}}{q-q^{-1}}%
~,\quad i,j=0,1,
\end{eqnarray}%
with $\mathcal{A}_{ij}=(-1)^{i+j}2$ being the Cartan matrix. In addition,
for $i\neq j$ the $q$-deformed Chevalley-Serre relations hold, 
\begin{equation}
x_{i}^{3}x_{j}-[3]_{q}x_{i}^{2}x_{j}x_{i}+[3]_{q}x_{i}x_{j}x_{i}^{2}-x_{j}x_{i}^{3}=0,\quad x=e,f\;.
\label{CS}
\end{equation}%
The coproduct of $U_{q}(\widehat{sl}_{2})$ is given by 
\begin{equation}
\Delta (e_{i})=1\otimes e_{i}+q^{h_{i}}\otimes e_{i},\quad \Delta
(f_{i})=f_{i}\otimes q^{-h_{i}}+1\otimes f_{i},\quad \Delta
(q^{h_{i}})=q^{h_{i}}\otimes q^{h_{i}}\;.  \label{cop}
\end{equation}%
The opposite coproduct $\Delta ^{\text{op}}$ is obtained by permuting the
two factors. There exists an universal $R$-matrix intertwining these two
coproduct structures 
\begin{equation}
\mathbf{R\,}\Delta (x)=\Delta ^{\text{op}}(x)\,\mathbf{R},\quad x\in U_{q}(%
\widehat{sl}_{2}),\quad \mathbf{R}\in U_{q}(b_{+})\otimes U_{q}(b_{-})\;.
\label{inter}
\end{equation}%
Here $U_{q}(b_{\pm })$ denote the upper and lower Borel subalgebra,
respectively.
\end{proposition}

Before we can define the six-vertex model in terms of representations of the
quantum group $U_{q}(\widehat{sl}_{2})$ we need one more ingredient, Jimbo's
evaluation homomorphism.

\begin{proposition}
Let $z\in \mathbb{C}$ be nonzero then the mapping 
\begin{equation}
e_{0}\rightarrow z\,f,\;f_{0}\rightarrow z^{-1}e,\;q^{h_{0}}\rightarrow
q^{-h},\;e_{1}\rightarrow e,\;f_{1}\rightarrow f,\;q^{h_{1}}\rightarrow
q^{h}\quad .  \label{ev}
\end{equation}%
defines an algebra homomorphism ev$_{z}:U_{q}(\widehat{sl}_{2})\rightarrow
U_{q}(sl_{2})$. Here $U_{q}(sl_{2})$ is isomorphic to the Hopf algebra
generated by either \{$e_{1},f_{1},q^{h_{1}}$\} or \{$e_{0},f_{0},q^{h_{0}}$%
\}.
\end{proposition}

Denote by $\pi _{z}^{(n)}=\pi ^{(n)}\circ ev_{z}:U_{q}(\widetilde{sl}%
_{2})\rightarrow \mbox{End}\mathbb{C}^{n+1}$ the spin $n/2$ evaluation
representation of the quantum group defined by the relations 
\begin{eqnarray}
\pi ^{(n)}(e)\left\vert m\right\rangle &=&[n-m+1]_{q}\left\vert
m-1\right\rangle ,  \notag \\
\pi ^{(n)}(f)\left\vert m\right\rangle &=&[m+1]_{q}\left\vert
m+1\right\rangle ,\quad \pi ^{(n)}(q^{h})\left\vert m\right\rangle
=q^{n-2m}\left\vert m\right\rangle \quad .  \label{pin}
\end{eqnarray}%
Here $\{\left\vert m\right\rangle \}_{m=0}^{n}$ denotes the standard
orthonormal basis in $\mathbb{C}^{n+1}$. For each integer $n\geq 0$ consider
the intertwiner of the tensor product $\pi _{z}^{(n)}\otimes \pi _{1}^{(1)}$%
, i.e. 
\begin{equation}
R^{(n+1)}(z)=\left( \pi _{z}^{(n)}\otimes \pi _{1}^{(1)}\right) \mathbf{R}%
\in \mbox{End}\left( \mathbb{C}^{n+1}\otimes \mathbb{C}^{2}\right) ,
\label{fusL}
\end{equation}%
whose matrix elements w.r.t. the second factor are calculated to be (up to
an overall normalizing factor) 
\begin{eqnarray}
\left\langle 0\right\vert R^{(n+1)}(z)\left\vert 0\right\rangle &=&zq\,\pi
^{(n)}(q^{h/2})-\pi ^{(n)}(q^{-h/2}),\quad  \notag \\
\left\langle 0\right\vert R^{(n+1)}(z)\left\vert 1\right\rangle
&=&zq\,(q-q^{-1})\pi ^{(n)}(q^{h/2})\pi ^{(n)}(f),  \notag \\
\left\langle 1\right\vert R^{(n+1)}(z)\left\vert 0\right\rangle
&=&(q-q^{-1})\pi ^{(n)}(e)\pi ^{(n)}(q^{-h/2}),\quad  \notag \\
\left\langle 1\right\vert R^{(n+1)}(z)\left\vert 1\right\rangle &=&zq\,\pi
^{(n)}(q^{-h/2})-\pi ^{(n)}(q^{h/2})\ .  \label{fusL2}
\end{eqnarray}%
We now introduce the six-vertex fusion hierarchy \cite{KR}.

\begin{definition}
Define the six-vertex fusion matrix $T^{(n)}:\left( \mathbb{C}^{2}\right)
^{\otimes M}\rightarrow \left( \mathbb{C}^{2}\right) ^{\otimes M}$ of degree 
$n\in \mathbb{N}$ by setting for some fixed integer $M\geq 1$ 
\begin{equation}
T^{(n)}(z)=\mbox{Tr}_{0}R_{0M}^{(n)}(zq^{n})\cdots R_{01}^{(n)}(zq^{n})\ .
\label{Tn}
\end{equation}%
The two special elements 
\begin{equation}
T^{(2)}(zq^{-2})\equiv T(z)\quad \quad \text{and}\quad \quad
T^{(1)}(z)\equiv (zq^{2}-1)^{M}~id\;,  \label{T}
\end{equation}%
are called the six-vertex row-to-row transfer matrix $T$ and the quantum
determinant \cite{KRS} $T^{(1)}$, respectively.
\end{definition}

\begin{proposition}[\protect\cite{KR}]
The fusion matrices satisfy the functional equation 
\begin{equation}
T^{(n)}(z)T^{(2)}(zq^{-2})=\left( zq^{2}-1\right)
^{M}T^{(n+1)}(zq^{-2})+\left( z-1\right) ^{M}T^{(n-1)}(zq^{2})  \label{fus}
\end{equation}%
and hence can be successively generated from the transfer matrix and quantum
determinant.
\end{proposition}

\begin{proof}
The assertion follows from the exact sequence \cite{CPbook}, 
\begin{equation}
0\rightarrow \pi _{w^{\prime }}^{(n-1)}\overset{\imath }{\hookrightarrow }%
\pi _{w}^{(n)}\otimes \pi _{z}^{(1)}\overset{p}{\rightarrow }\pi _{w^{\prime
\prime }}^{(n+1)}\rightarrow 0,\quad w=w^{\prime }q^{-1}=w^{\prime \prime
}q=zq^{n+1}  \label{fusseq}
\end{equation}%
See also \cite{KQ3} for details such as the explicit form of the inclusion $%
\imath $ and projection map $p$ needed to derive (\ref{fus}).
\end{proof}

The statistical mechanics model is now defined in terms of the partition
function $Z_{\text{6v}}=\mbox{Tr}T^{M^{\prime }}$ which is the physical
quantity of interest. Here $M$ is the number of lattice columns and $%
M^{\prime }$ the number of lattice rows. In this definition we have imposed
periodic boundary conditions on the lattice. This model is said to be
integrable as the transfer matrix $T$ has infinitely many conserved
quantities.

\begin{corollary}
From the definition (\ref{Tn}) and the existence of the universal $R$-matrix
it immediately follows that 
\begin{equation}
\lbrack T^{(m)}(z),T^{(n)}(w)]=0\;,\qquad \forall z,w\in \mathbb{C}\text{,\ }%
m,n\in \mathbb{N}\ .  \label{symm}
\end{equation}

\begin{proof}
The assertion is a direct consequence of the Yang-Baxter equation, $\mathbf{R%
}_{12}\mathbf{R}_{13}\mathbf{R}_{23}=\mathbf{R}_{23}\mathbf{R}_{13}\mathbf{R}%
_{12}$.
\end{proof}
\end{corollary}

For later purposes we note further symmetries of the six-vertex fusion
matrices. Their proof follows by direct calculation.

\begin{lemma}
Let $\sigma ^{x},\sigma ^{y},\sigma ^{z}$ denote the Pauli matrices. Define
the following matrices acting on $\left( \mathbb{C}^{2}\right) ^{\otimes M}$%
, 
\begin{equation}
S^{z}=\frac{1}{2}\sum_{m=1}^{M}\sigma _{m}^{z},\text{\quad }\mathfrak{R}%
=\prod\limits_{m=1}^{M}\sigma _{m}^{x},\text{\quad }\mathfrak{S}%
=\prod\limits_{m=1}^{M}\sigma _{m}^{z}\;.  \label{SzRS}
\end{equation}%
Then for any integer $n\geq 1$ the fusion matrix $T^{(n)}$ commutes with $%
S^{z},\mathfrak{R}$ and $\mathfrak{S}$.
\end{lemma}

Throughout this article we will implicitly assume that we always work in a
basis of eigenvectors of the total-spin operator $S^{z}$, i.e. we will treat 
$S^{z}$ as a diagonal matrix with half-integer entries in the interval $%
[-M/2,M/2]$.

\begin{lemma}
The transfer matrix satisfies the identity%
\begin{equation}
T(z,q)=T(zq^{2},q^{-1})^{t}  \label{transp}
\end{equation}%
and hence is normal provided $|q|=1$; see (\ref{symm}).
\end{lemma}

\section{The Auxiliary Matrix}

Henceforth we will specialize to the case when $q$ is a primitive root of
unity of order $N\geq 3$. The six-vertex fusion hierarchy stays well-defined
in this limit. We set $N^{\prime }=N$ if the order is odd and $N^{\prime
}=N/2$ if it is even.

\begin{proposition}
For any $r,s,z\in \mathbb{C}$ the following defines an $N^{\prime }$%
-dimensional representation $\pi ^{+}=\pi ^{+}(z;r,s)$ of the upper Borel
subalgebra $U_{q}(b_{+})$: 
\begin{eqnarray}
\pi ^{+}(q^{h_{1}})\left\vert n\right\rangle &=&\pi
^{+}(q^{-h_{0}})\left\vert n\right\rangle =rq^{-2n}\left\vert n\right\rangle
,  \notag \\
\pi ^{+}(e_{0})\left\vert n\right\rangle &=&z\left\vert n+1\right\rangle
,\quad \quad \pi ^{+}(e_{0})\left\vert N^{\prime }-1\right\rangle =0,  \notag
\\
\pi ^{+}(e_{1})\left\vert n\right\rangle &=&\frac{s+1-q^{2n}-sq^{-2n}}{%
(q-q^{-1})^{2}}\;\left\vert n-1\right\rangle ,\quad \pi
^{+}(e_{1})\left\vert 0\right\rangle =0\quad .  \label{piplus}
\end{eqnarray}%
Here $\{\left\vert n\right\rangle \}_{n=0}^{N^{\prime }-1}$ denotes the
standard orthonormal basis in $\mathbb{C}^{N^{\prime }}$. Let $\omega $
denote the algebra automorphism $\{e_{1},e_{0},q^{h_{1}},q^{h_{0}}\}%
\rightarrow \{e_{0},e_{1},q^{h_{0}},q^{h_{1}}\}$ and set 
\begin{equation}
\pi ^{-}:=\pi ^{+}\circ \omega \;.  \label{piminus}
\end{equation}
\end{proposition}

A short calculation shows that the defining relations of the quantum group
are satisfied in the representations $\pi ^{\pm }$. In order to make contact
with previous results in the literature we have the following

\begin{remark}
Note that the representation $\pi ^{-}$ is a particular root-of-unity
restriction of the representation in \cite{RW02}. For the choice of
parameters $r=\mu ^{-1}q^{-1}$ and $s=\mu ^{-2}$ the representation (\ref%
{piplus}) coincides with the representation used in \cite{KQ,KQ2,KQ3,KQ4}.
\end{remark}

Let the matrix $L=\alpha \otimes \sigma ^{+}\sigma ^{-}+\beta \otimes \sigma
^{+}+\gamma \otimes \sigma ^{-}+\delta \otimes \sigma ^{-}\sigma ^{+}$ be
the intertwiner of the tensor product $\pi ^{+}\otimes \pi _{z=1}^{(1)}$ of
evaluation representations, explicitly 
\begin{equation*}
L\,(\pi ^{+}\otimes \pi _{1}^{(1)})\Delta (x)=\left( (\pi ^{+}\otimes \pi
_{1}^{(1)})\Delta ^{\text{op}}(x)\right) L,\quad \forall x\in U_{q}(b^{+})\ .
\end{equation*}
From this relation the matrix elements up to an overall normalization factor
are computed to 
\begin{eqnarray}
\alpha &=&zs/r\,\pi ^{+}(q^{\frac{h_{1}}{2}})-\pi ^{+}(q^{-\frac{h_{1}}{2}%
}),\quad \beta =(q-q^{-1})\pi ^{+}(e_{0})\pi ^{+}(q^{-\frac{h_{0}}{2}}),\; 
\notag \\
\gamma &=&\left( q-q^{-1}\right) \pi ^{+}(e_{1})\pi ^{+}(q^{-\frac{h_{1}}{2}%
}),\quad \delta =z\,rq^{2}\,\pi ^{+}(q^{-\frac{h_{1}}{2}})-\pi ^{+}(q^{\frac{%
h_{1}}{2}})\ .  \label{L}
\end{eqnarray}%
Notice that (up to a simple gauge transformation) the intertwiner for the
representation $\pi ^{-}$ is obtained via the permutation $\{\alpha ,\beta
,\gamma ,\delta \}\rightarrow \{\delta ,\gamma ,\beta ,\alpha \}$ which
corresponds to spin reversal, i.e. $L\rightarrow (1\otimes \sigma
^{x})L(1\otimes \sigma ^{x})$.

\begin{definition}
Define the auxiliary matrix in terms of the intertwiner $L$ as the trace of
the following operator product, 
\begin{equation}
Q(z;r,s)=\mbox{Tr}_{0}L_{0M}(z;r,s)\cdots L_{01}(z;r,s)\ .  \label{Q}
\end{equation}
\end{definition}

This matrix commutes by construction with the fusion matrices as the
intertwiner $L$ must satisfy the Yang-Baxter equation \cite{DJMM90}%
\begin{equation*}
L_{12}(z/w)L_{13}(z)R_{23}^{(2)}(w)=R_{23}^{(2)}(w)L_{13}(z)L_{12}(z/w)~.
\end{equation*}
Furthermore, $Q$ preserves two of the symmetries (\ref{symm}) \cite{KQ,KQ2},
i.e. 
\begin{equation}
\lbrack Q(z;r,s),T^{(n)}(w)]=[Q(z;r,s),S^{z}]=[Q(z;r,s),\mathfrak{S}]=0\ .
\label{QT0}
\end{equation}%
Spin-reversal symmetry on the other hand is broken \cite{KQ,KQ2}. To see
this we first note that dependence of the auxiliary matrix $Q$ on the
parameter $r$ can be easily extracted.

\begin{lemma}
The auxiliary matrix can be simplified as follows 
\begin{equation}
Q(z;r,s)=r^{-S^{z}}Q(z;1,s)\equiv r^{-S^{z}}Q(z;s)\;.  \label{Qsimp}
\end{equation}
\end{lemma}

\begin{proof}
Because the auxiliary matrix preserves the total spin a general matrix
element $\left\langle \sigma _{1},...,\sigma _{M}|Q|\varrho _{1},...,\varrho
_{M}\right\rangle $ must contain as many $\beta $-matrices as it does $%
\gamma $-matrices whose $r$-dependence cancels against each other. Each
occurrence of an $\alpha $-matrix contributes a factor $r^{-1/2}$ while each 
$\delta $-matrix gives rise to a factor $r^{1/2}$.
\end{proof}

As an immediate consequence of the decomposition of the auxiliary matrix
w.r.t. to the parameters $r,s$ we have the following

\begin{corollary}
The auxiliary matrix constructed in this article can be identified with the
auxiliary matrix in \cite{KQ,KQ2,KQ4} via the relation 
\begin{equation}
Q_{\mu }(z)=\mu ^{S^{z}}q^{S^{z}}Q(z;\mu ^{-2})\ .  \label{identi}
\end{equation}
\end{corollary}

We now easily find the identities

\begin{lemma}
Under spin reversal the auxiliary matrix transforms as 
\begin{eqnarray}
\mathfrak{R}Q(z,q;s)\mathfrak{R} &=&(-z)^{M}q^{M+2S^{z}}s^{\frac{M}{2}%
-S^{z}}Q(z^{-1}q^{-2}s^{-1},q;s)^{t}  \notag \\
&=&Q(zq^{2}s,q^{-1};s^{-1})^{t}=q^{2S^{z}}Q(zs,q;s^{-1})\ .  \label{RQ}
\end{eqnarray}%
From the last equation the conjugate transpose of the auxiliary matrix is
deduced to be 
\begin{equation}
Q(z,q;s)^{\ast }=Q(\bar{z},q^{-1};\bar{s})^{t}=q^{2S^{z}}Q(\bar{z}q^{-2},q;%
\bar{s})\;.  \label{Qct}
\end{equation}
\end{lemma}

\begin{proof}
The transformation properties (\ref{RQ}) are a direct consequence of the
identities in \cite{KQ,KQ2,KQ4} and (\ref{identi}).
\end{proof}

As we want to solve the eigenvalue problem of the six-vertex transfer matrix
in terms of the auxiliary matrix we need to show that $Q$ is diagonalizable.
Further, we wish to establish the analytic properties of the eigenvalues.

\begin{claim}
Let $z,w,s,t\in \mathbb{C}$ be arbitrary independent complex numbers. Then
the following commutation relation holds for all primitive roots of unity 
\begin{equation}
\lbrack Q(z;s),Q(w;t)]=0\ .  \label{Qcomm}
\end{equation}
\end{claim}

\begin{proof}[Proof for $N=3,4,6$.]
A sufficient (not necessary) condition for the above commutation relation to
hold is to show that the intertwiner, say $S$, for the tensor product $\pi
^{+}(z;1,s)\otimes \pi ^{+}(w;1,t)$ exists. If it does, $S$ is bound to
satisfy the Yang-Baxter equation \cite{DJMM90} $%
S_{12}L_{13}(z;s)L_{23}(w;t)=L_{23}(w;t)L_{13}(z;s)S_{12}$ from which the
assertion immediately follows. Unlike the case when $q$ is not a root of
unity the existence of such an intertwiner cannot be deduced from the
existence of the universal R-matrix; see \cite{CPbook} for details and \cite%
{KQ} for the connection with the present discussion. It appears that at the
moment the only way known to prove existence of $S$ is through explicit
construction. This has been carried out for the case $N=3$ in \cite{KQ2}.
Here we state the intertwiners for $N=4,6$ in the appendix of this article.
Numerical checks carried out for $N=5,7,8$ and $M\leq 11$ confirm (\ref%
{Qcomm}) also in these cases.
\end{proof}

Henceforth, we take (\ref{Qcomm}) as a working hypothesis. The two immediate
implications are:

\noindent 1. According to (\ref{Qct}) the auxiliary matrix is normal, $%
[Q(z;s),Q(z;s)^{\ast }]=0$, and hence diagonalizable.\medskip

\noindent 2. The eigenvalues of $Q(z;s)$ are polynomials in $z$ whose degree
is at most $M$. By construction we know that (\ref{Q}) can be decomposed as
follows, 
\begin{equation*}
Q(z,q;s)=\sum_{m=0}^{M}Q_{m}(s,q)z^{m},
\end{equation*}%
and (\ref{Qcomm}) then ensures that the coefficients $\{Q_{m}(s,q)\}$ can be
simultaneously diagonalized. In addition we infer from (\ref{RQ}) that the
coefficients are related via%
\begin{equation}
Q_{M-m}(s,q)=(-1)^{M}q^{M-2S^{z}}s^{\frac{M}{2}+S^{z}}Q_{m}(s^{-1},q^{-1})\;.
\label{coeff}
\end{equation}%
Finally, we can make use of the fact that the intertwiner (\ref{L}) is a
lower-triangular matrix at $z=0$ from which we deduce the following identity
for the zeroth term of the polynomial, 
\begin{equation}
Q(0;s)=Q_{m=0}(s,q)=(-1)^{M}\mbox{Tr}_{\pi
^{+}}q^{-h_{1}S^{z}}=(-1)^{M}\sum_{\ell =0}^{N^{\prime }-1}q^{2\ell S^{z}}\;.
\label{norm}
\end{equation}%
Note that because of (\ref{coeff}) this expression determines at the same
time whether $Q$ has the maximal degree $M$ or not.

\subsection{The limit $s\rightarrow 0$ of the auxiliary matrix}

For later purposes we end this section by introducing two special limits of
the auxiliary matrix.

\begin{definition}
Let $\pi ^{\pm }=\pi ^{\pm }(z;1,s)$ be the representations (\ref{piplus}), (%
\ref{piminus}) and $L^{\pm }$ the respective intertwiners (\ref{L}). Then
the limit $s\rightarrow 0$ is well-defined and we set 
\begin{equation}
Q^{\pm }(z):=\lim_{s\rightarrow 0}\mbox{Tr}_{\pi ^{\pm }}L_{0M}^{\pm
}(z)\cdots L_{01}^{\pm }(z)\;.  \label{Qpm}
\end{equation}
\end{definition}

From the remarks after equation (\ref{L}) it follows immediately that the
two operators are related via spin-reversal $\mathfrak{R}Q^{-}=Q^{+}%
\mathfrak{R}$. The motivation for the above definition will become apparent
in Section 5.

\section{Baxter's TQ Equation}

We are now ready to relate our construction of the auxiliary matrix to the
six-vertex fusion hierarchy. The link is provided by a functional equation
similar to Baxter's famous $TQ$ equation. In contrast to his approach we can
now exploit the representation theory of quantum groups in its derivation.

\begin{proposition}
The auxiliary matrix (\ref{Q}) and the six-vertex transfer matrix (\ref{T})
obey the following operator $q$-difference equation for any primitive root
of unity $q$, 
\begin{equation}
Q(z;s)T(z)=(z-1)^{M}q^{S^{z}}Q(zq^{2};sq^{-2})+(zq^{2}-1)^{M}q^{-S^{z}}Q(zq^{-2};sq^{2})\;.
\label{TQ0}
\end{equation}
Here $S^{z}$ is the total-spin operator defined in (\ref{SzRS}).
\end{proposition}

\begin{proof}
The strategy of the proof follows closely the one in \cite{KQ} by means of
the identification (\ref{identi}). For the derivation of (\ref{TQ0}) it is
more convenient to re-introduce the parameter $r$ of the representation (\ref%
{piplus}) and then at the end use (\ref{Qsimp}) for taking the limit $%
r\rightarrow 1$. One verifies the following non-split exact sequence of
representations of $U_{q}(b_{+})$, 
\begin{equation*}
0\rightarrow \pi ^{+}(zq^{2};rq^{-1},sq^{-2})\overset{\imath }{%
\hookrightarrow }\pi ^{+}(z;r,s)\otimes \pi _{z}^{(1)}\overset{p}{%
\rightarrow }\pi ^{+}(zq^{-2};rq,sq^{2})\rightarrow 0\;.
\end{equation*}%
This determines the functional equation by the same arguments as presented
in \cite{KQ}.
\end{proof}

Notice that the above functional relation (\ref{TQ0}) differs from Baxter's
original $TQ$ equation as his matrix does not depend on extra parameters
which shift. However, as a trivial consequence we now deduce that the two
auxiliary matrices (\ref{Qpm}) obey a functional equation which is of
Baxter's type.

\begin{corollary}[Baxter's TQ-equation]
In the limit $s\rightarrow 0$ we have 
\begin{equation}
Q^{\pm }(z)T(z)=(z-1)^{M}q^{\pm S^{z}}Q^{\pm }(zq^{2})+(zq^{2}-1)^{M}q^{\mp
S^{z}}Q^{\pm }(zq^{-2})\;.  \label{TQpm}
\end{equation}%
Notice that the equation for $Q^{-}$ follows from the one for $Q^{+}$ by
spin-reversal.
\end{corollary}

Since transfer and auxiliary matrix are both simultaneously diagonalizable
we can interpret the functional equations (\ref{TQ0}), (\ref{TQpm}) on the
level of eigenvalues.

\begin{corollary}
Let $Q(z;s)$ be an eigenvalue of the auxiliary matrix (\ref{Q}) which is not
identical zero. (We denote eigenvalues and operators by the same symbol.)
Then the corresponding eigenvalue of the fusion matrix of degree $n$ is
given by the formula 
\begin{equation}
T^{(n)}(z)=q^{\pm (n+1)S^{z}}Q^{\pm }(z)Q^{\pm }(zq^{2n})\sum_{\ell =1}^{n}%
\frac{q^{\mp 2\ell S^{z}}(zq^{2\ell }-1)^{M}}{Q^{\pm }(zq^{2\ell })Q^{\pm
}(zq^{2\ell -2})}\;.  \label{specTn}
\end{equation}
\end{corollary}

\begin{proof}
From (\ref{fus}) the spectrum of the fusion matrices (\ref{Tn}) is
calculated via induction.
\end{proof}

\section{Another Functional Equation}

To fully analyze the structure of the eigenvalues of the auxiliary matrix
Baxter's $TQ$-equation is not sufficient. Instead we make use of another
functional equation which by the same philosophy as before follows from the
decomposition of a tensor product of representations.

\begin{proposition}
The auxiliary matrix (\ref{Q}) obeys the functional equation%
\begin{equation}
Q(zq^{2}/s;s)Q(z;t)=Q(zq^{2}/s;stq^{-2})\left[ (zq^{2}-1)^{M}+q^{N^{\prime
}(M-S^{z})}T^{(N^{\prime }-1)}(zq^{2})\right]  \label{func}
\end{equation}%
for arbitrary complex numbers $s,t\in \mathbb{C}$.
\end{proposition}

\begin{proof}
The proof follows the same strategy as the one for the functional equation (%
\ref{TQ0}). We invoke the identity (\ref{identi}) and refer the reader to 
\cite{KQ4} for the derivation of the exact sequence%
\begin{multline*}
0\rightarrow \pi ^{+}(z^{\prime };s^{\prime \frac{1}{2}}q^{-1},s^{\prime
})\rightarrow \pi ^{+}(zq^{2}/s;s^{\frac{1}{2}}q^{-1},s)\otimes \pi
^{+}(z;t^{\frac{1}{2}}q^{-1},t)\rightarrow \\
\pi ^{+}(z^{\prime \prime };s^{\prime \prime }{}^{\frac{1}{2}%
}q^{-1},s^{\prime \prime })\otimes \pi _{zq^{N^{\prime }+1}}^{(N^{\prime
}-2)}\rightarrow 0
\end{multline*}%
where with regard to equation (51) in \cite{KQ4} we have set%
\begin{equation*}
s=\mu ^{-2},\;t=\nu ^{-2},\;s^{\prime \frac{1}{2}}=(\mu \nu
q)^{-1},\;s^{\prime \prime \frac{1}{2}}=(\mu \nu q^{-N^{\prime
}+1})^{-1},\;z^{\prime }=z^{\prime \prime }=zq^{2}/s\ .
\end{equation*}%
The remaining proof now follows the analogous steps as in \cite{KQ4}, see
the appendix therein, employing at the end (\ref{identi}) to arrive at (\ref%
{func}).
\end{proof}

\begin{corollary}
Let $M\in 2\mathbb{N}+1$ and $N\in 2\mathbb{N}$. Then all the eigenvalues of
the auxiliary matrix are of the form%
\begin{equation}
Q(z;s)=-\frac{1-q^{2N^{\prime }S^{z}}}{1-q^{2S^{z}}}\prod_{j=1}^{n}(1-z%
\,x_{j}^{+})\prod_{j=1}^{M-n}(1-zs\,x_{j}^{-}),\quad n=\frac{M}{2}-S^{z},
\label{Qpoly}
\end{equation}%
where the parameters $x_{j}^{\pm }$ do only depend on $q$. This implies the
following decomposition of the auxiliary matrix, $%
Q(z;s)=Q^{-}(0)^{-1}~Q^{+}(z)Q^{-}(zs)$. Moreover, we have the identity%
\begin{equation}
Q^{-}(z)=q^{N^{\prime }M}Q^{-}(0)Q^{+}(z)\sum_{\ell =1}^{N^{\prime }}\frac{%
q^{-2\ell S^{z}}(zq^{2\ell }-1)^{M}}{Q^{+}(zq^{2\ell })Q^{+}(zq^{2\ell -2})}%
\ .  \label{sumQp}
\end{equation}
\end{corollary}

\begin{proof}
We state a slightly simpler version of the proof given in \cite{KQ4} which
applies to the case of arbitrary positive integers $N,M$. As $N$ is even we
have $q^{N^{\prime }}=-1$ and since $M$ is odd the eigenvalues of the
total-spin are half-odd integers, i.e. $S^{z}=-M/2,-M/2+1,...,M/2$. Thus, we
can conclude from (\ref{norm}) and (\ref{coeff}) that the eigenvalues of the
auxiliary matrix are polynomials in the spectral variable $z$ of degree $M$
and are non-vanishing at the origin. Hence, they can be written in the form%
\begin{equation*}
Q(z;s)=\mathcal{N}\prod_{j=1}^{n}(1-z\,x_{j}^{+})\prod_{j=1}^{M-n}(1-z%
\,y_{j}(s)),\quad \mathcal{N}=-\frac{1-q^{2N^{\prime }S^{z}}}{1-q^{2S^{z}}}\
.
\end{equation*}%
where the normalization constant $\mathcal{N}=Q^{+}(0)=q^{-2S^{z}}Q^{-}(0)$
is given by (\ref{norm}), (\ref{RQ}) and (\ref{Qpm}). Here we have split the
zeroes of the eigenvalues into two groups: the zeroes $x_{i}^{+}$ do not
depend on the parameter $s$, while the zeroes $y_{i}=y_{i}(s)$ do depend on
it. We allow for the extreme cases $n=0$ and $n=M$. Employing the functional
equation (\ref{func}) we find that the ratio%
\begin{equation*}
\frac{Q(zq^{2}/s;s)Q(z;t)}{Q(zq^{2}/s;stq^{-2})}=\mathcal{N}%
\prod_{j=1}^{n}(1-z\,x_{j}^{+})\prod_{j=1}^{M-n}\frac{%
(1-zq^{2}y_{j}(s)/s)(1-zy_{j}(t))}{1-zq^{2}y_{j}(stq^{-2})/s}
\end{equation*}%
must be independent of the arbitrary complex parameters $s,t$ which implies
that the zeroes $y_{i}$ depend linearly on $s$, i.e. $y_{i}(s)=x_{i}^{-}s$
for some $x_{i}^{-}$ which only depends on $q$. Invoking now once more the
transformations (\ref{RQ}) to determine $n$ this proves the first assertion.
The second now trivially follows also from (\ref{RQ}) and (\ref{Qpm}). The
third and last identity is immediate from the functional equation (\ref{func}%
) and the expression (\ref{specTn}) for the spectrum of the fusion matrices.
\end{proof}

\begin{remark}
The general case is slightly more involved. It can then happen that some
eigenvalues of the auxiliary matrix are identical zero or vanish at the
origin. The relation $n=M/2-S^{z}$ also ceases to be valid. Particularly
noteworthy is the fact that some zeroes might organize in strings ($%
x_{i}^{\pm },x_{i}^{\pm }q^{2},...,x_{i}^{\pm }q^{2N^{\prime }-2}$)
signaling degeneracies in the spectrum of the transfer matrix. We refer the
reader to \cite{KQ4} for details and further references.
\end{remark}

\begin{remark}
Note that the identity is the six-vertex analogue of a functional relation
which has been conjectured in \cite{FM8v, FM8v2} for the eight-vertex case.
See the conclusion of \cite{KQ4} for further comments and references. The
matrix $Q^{+}$ is the six-vertex analogue of Baxter's $Q$-operator in \cite%
{Bx72} and $Q^{-}$ corresponds to its counterpart under spin-reversal,
compare with the discussion in \cite{FM8v3}. While this eight-vertex result
implies the existence of the six-vertex solutions $Q^{\pm },$ the algebraic
form of Baxter's $Q$-operator \cite{Bx72} does not allow one to take
directly the trigonometric limit and obtain an explicit six-vertex $Q$%
-operator.
\end{remark}

\section{Quantum Wronskian}

In this final section we now wish to make contact with the Bethe ansatz
analysis of the six-vertex model \cite{Lieb67,St67} by deriving from the
results on the spectra of the auxiliary matrices (\ref{Q}), (\ref{Qpm}) the
Bethe ansatz equations. In fact, we will actually do more than that: the
existence of two linearly independent solutions to Baxter's $TQ$-equation
together with (\ref{sumQp}) implies a functional equation which leads to two
sets of Bethe ansatz equations, one above and one below the equator.

\begin{corollary}
Let $M$ be odd and $N$ even as before. Renormalizing%
\begin{equation*}
\mathcal{Q}^{\pm }(z)=Q^{\pm }(0)^{-1}Q^{\pm }(z)
\end{equation*}%
the two solutions (\ref{Qpm}) of Baxter's $TQ$-equation then satisfy for any 
$1\leq n\leq N^{\prime }$%
\begin{equation}
T^{(n)}(z)=-\frac{q^{nS^{z}}\mathcal{Q}^{+}(zq^{2n})\mathcal{Q}%
^{-}(z)-q^{-nS^{z}}\mathcal{Q}^{+}(z)\mathcal{Q}^{-}(zq^{2n})}{%
q^{S^{z}}-q^{-S^{z}}}\ .  \label{wronski}
\end{equation}
\end{corollary}

\begin{proof}
The identities follow by a simple calculation from (\ref{sumQp}) and (\ref%
{specTn}).
\end{proof}

The case $n=1$ of (\ref{wronski}) is special as it only involves the
eigenvalues of the auxiliary matrices (\ref{Qpm}) and the quantum
determinant (\ref{T}), which is explicitly known. Hence, this equation,
which plays the analogous role of a Wronskian, is sufficient to solve the
eigenvalue problem of the fusion hierarchy. To stress this point we state
equation (\ref{wronski}) for $n=1$ in a different form.

\begin{corollary}
For $M$ odd and $N$ even denote by%
\begin{equation*}
e_{k}^{\pm }=e_{k}(x_{1}^{\pm }q^{-1},...,x_{n_{\pm }}^{\pm }q^{-1}),\quad
k=0,1,...,\frac{M}{2}\mp S^{z}
\end{equation*}%
the elementary symmetric polynomials in the zeroes of the eigenvalues of $%
Q^{\pm }$; see (\ref{Qpoly}). We use the convention $e_{0}^{\pm }=1$ and $%
e_{k}^{\pm }=0$ for $k>M/2\mp S^{z}$. Then for any integer $0\leq m\leq M$
we have%
\begin{equation}
\binom{M}{m}=\sum_{k=0}^{m}\frac{q^{S^{z}-m+2k}-q^{-S^{z}+m-2k}}{%
q^{S^{z}}-q^{-S^{z}}}\;e_{k}^{+}e_{m-k}^{-}\ .  \label{wronski2}
\end{equation}%
Notice that these are $M$ quadratic equations in the $M$ variables $%
\{e_{k}^{\pm }\}_{k=1}^{M/2\mp S^{z}}$.
\end{corollary}

\begin{remark}
Notice that (\ref{wronski2}) only holds true for an odd number of lattice
columns $M$. While we have derived it here for even roots of unity only, it
extends to the case when $q$ is of odd order as well. However, then the
number of solutions drastically decreases and is in general strictly less
than the maximal number $\binom{M}{M/2-S^{z}}$ given by the dimension of a
fixed spin sector. For example, when $q$ is a root of unity of order three
there exists only one solution to (\ref{wronski2}); see \cite{KQ4} and
references therein for details.

The equations (\ref{wronski2}) do also apply for odd $M$ and $q$ being not a
root of unity. However, then the representation (\ref{piplus}) defining the
auxiliary matrix must be chosen infinite-dimensional \cite{RW02,KQ3}.
\end{remark}

\begin{corollary}
Parametrizing the zeroes $x_{i}^{\pm }$ in (\ref{Qpoly}) as%
\begin{equation}
x_{i}^{\pm }q^{-1}=\frac{\sin \frac{\pi }{2}\left( k_{i}^{\pm }+2n/N\right) 
}{\sin \frac{\pi }{2}\left( k_{i}^{\pm }-2n/N\right) }=\frac{e^{i\pi
k_{i}^{\pm }}q-1}{e^{i\pi k_{i}^{\pm }}-q},\quad q=e^{2\pi in/N}\ 
\label{ki}
\end{equation}%
and setting $B_{ij}^{\pm }=1-2\cos (\frac{2\pi n}{N})e^{i\pi k_{j}^{\pm
}}+e^{i\pi (k_{i}^{\pm }+k_{j}^{\pm })}$ we have the identities,%
\begin{equation}
e^{i\pi Mk_{i}^{\pm }}\prod\limits_{j=1}^{n_{\pm }}B_{ij}=(-)^{n_{\pm
}-1}\prod\limits_{j=1}^{n_{\pm }}B_{ji},\quad n_{\pm }=\frac{M}{2}\mp S^{z}\
,  \label{BAE}
\end{equation}%
which are called the six-vertex Bethe ansatz equations above and below the
equator. In contrast to the equations (\ref{wronski2}) the Bethe ansatz
equations (\ref{BAE}) are polynomial equations of order $M$ in $M/2\mp S^{z}$
variables.
\end{corollary}

\begin{proof}
Evaluate equation (\ref{wronski}) at $n=1$ and $z=1/x_{i}^{\pm
}q,1/x_{i}^{\pm }q^{-1}$.
\end{proof}

\section{Conclusions}

Using representation theory of quantum groups we have derived the Bethe
ansatz equations of the six-vertex model at even roots of unity and at odd
numbers of lattice columns $M$. Using the concept of auxiliary matrices we
have shown by explicit construction that there are two sets of Bethe ansatz
equations and that their solutions exist. Moreover, we proved that \emph{all}
the eigenvalues of the fusion hierarchy are determined by their solutions.
This follows from the fact that for even roots of unity and odd $M$ the
eigenvalues of the auxiliary matrix (\ref{Q}) are always non-vanishing
polynomials of degree $M$ which must satisfy Baxter's $TQ$-equation. This
establishes existence and completeness, albeit we have not shown that all
eigenvalues of the auxiliary matrix are \emph{different}, i.e. that there
are precisely $\binom{M}{M/2-S^{z}}$ different solutions to (\ref{BAE})
respectively (\ref{wronski2}). Our derivation rests on the crucial
commutation relations (\ref{Qcomm}) which we proved only for $N=4,6$ in this
article. Nevertheless, in our approach we were able to reduce the question
of existence and completeness to a well-defined representation theoretic
problem, namely the existence of an intertwiner for the tensor product of
two specific quantum group representations.

We conclude by emphasizing again the concrete practical implications of the
existence of two linearly independent solutions to the TQ equation. As we
derived in the text the eigenvalues of $Q^{\pm }$ satisfy for the identity (%
\ref{wronski}) which for $n=1$ leads to a \emph{quadratic} equation in $M$
variables with $M$ being the number of lattice columns; see equation (\ref%
{wronski2}). In contrast, the Bethe ansatz equations (\ref{BAE}) are
polynomial equations of order $M$ in $M/2$ variables. For numerical
investigations, often performed in physical applications, this reduction of
the polynomial order of the equations is crucial in particular when the
system size becomes large $M\gg 1$.\medskip

\noindent \textsc{Acknowledgement}. The author would like to thank the
organizers of the conference "Non-commutative Geometry and Representation
Theory in Mathematical Physics", Karlstads Universitet, Sweden, July 2004
and B.M. McCoy for comments. This work has been financially supported by the
EPSRC Grant GR/R93773/01 and a University Research Fellowship of the Royal
Society.

\appendix

\section{The intertwiners for $\protect\pi ^{+}(z;s)\otimes \protect\pi %
^{+}(w;t)$ when $N=4,6$}

In this section the intertwiners for the tensor product $\pi
^{+}(z;s)\otimes \pi ^{+}(w;t)$ of evaluation representations are stated for
roots of unity of order $N=4,6$. The construction employs the defining
equation of the intertwiner $S$ which is given by 
\begin{equation}
S[\pi ^{+}(z;s)\otimes \pi ^{+}(w;t)\Delta (x)]=\left[ (\pi ^{+}(z;s)\otimes
\pi ^{+}(w;t)\Delta ^{\text{op}}(x)\right] S,\;x\in U_{q}(b_{+})\;.
\label{S}
\end{equation}%
Without loss of generality we can set $w=1$ as it follows from the algebraic
structure of the equations (\ref{S}) that the intertwiner $S$ only depends
on the ratio of the spectral parameters, i.e. $S=S(z/w)$.

\subsection{The case $N=4$}

For roots of unity of order four the representation $\pi ^{+}(z;s)$ defining
the auxiliary matrix is two-dimensional and the equations (\ref{S}) are
easily solved for the Chevalley-Serre generators. One obtains the result 
\begin{equation*}
S=\left( 
\begin{array}{cccc}
1 & 0 & 0 & 0 \\ 
0 & \frac{1+t-(1+s)\lambda z}{(1+s)(1+z)} & \frac{1+\lambda }{1+z} & 0 \\ 
0 & \frac{1+\lambda }{1+z}~z & \frac{(1+s)z-(1+t)\lambda }{(1+s)(1+z)} & 0
\\ 
0 & 0 & 0 & \lambda%
\end{array}%
\right) \;.
\end{equation*}%
Here $\lambda $ is a free parameter and the matrix is computed with respect
to the basis $\{\left\vert 0,0\right\rangle ,\left\vert 0,1\right\rangle
,\left\vert 1,0\right\rangle ,\left\vert 1,1\right\rangle \}$ in the tensor
product $\pi ^{+}(z;s)\otimes \pi ^{+}(1;t)$. The arbitrary overall
normalization factor of $S$ is fixed by setting the matrix element of the
highest weight vector equal to one.

\subsection{The case $N=6$}

Now the representations $\pi ^{+}(z;s),\pi ^{+}(w=1;t)$ are each
three-dimensional. As $S$ commutes with the Cartan generators we immediately
deduce that $S$ must be block-diagonal w.r.t. the following decomposition of
the tensor product space, $\pi ^{+}(z;s)\otimes \pi ^{+}(1;t)=V_{1}\oplus
V_{2}\oplus V_{3}$ where the respective subspaces are spanned by the
following basis vectors 
\begin{eqnarray*}
V_{1} &=&\mbox{span}\{\left\vert 0,0\right\rangle ,\left\vert
1,2\right\rangle ,\left\vert 2,1\right\rangle \}, \\
V_{2} &=&\mbox{span}\{\left\vert 0,1\right\rangle ,\left\vert
1,0\right\rangle ,\left\vert 2,2\right\rangle \}, \\
V_{3} &=&\mbox{span}\{\left\vert 0,2\right\rangle ,\left\vert
1,1\right\rangle ,\left\vert 2,0\right\rangle \}\;.
\end{eqnarray*}%
The calculation is cumbersome but straightforward and one finds the
following solution up to a common normalization factor, 
\begin{eqnarray*}
S|_{V_{1}} &=&\left( 
\begin{array}{ccc}
1 & 0 & 0 \\ 
0 & \frac{(1-z)(zq^{-1}+t)}{(sz+q)(sz+q^{-1})} & \frac{(q+s)(t+zq^{-1})}{%
(sz+q)(sz+q^{-1})} \\ 
0 & \frac{z(zq+t)(t+q)}{(sz+q)(sz+q^{-1})} & \frac{(sz-t)(z+tq)}{%
(sz+q)(sz+q^{-1})}%
\end{array}%
\right) , \\
S|_{V_{2}} &=&\left( 
\begin{array}{ccc}
\frac{(1-z)(szq^{-1}+1)}{(sz+q)(sz+q^{-1})} & \frac{(sq+1)(1+szq^{-1})}{%
(sz+q)(sz+q^{-1})} & 0 \\ 
\frac{z(sz+q)(t+q^{-1})}{(sz+q)(sz+q^{-1})} & \frac{(sz-t)(sz+q)}{%
(sz+q)(sz+q^{-1})} & 0 \\ 
0 & 0 & \frac{(zq+t)(zq^{-1}+t)}{(sz+q)(sz+q^{-1})}%
\end{array}%
\right) , \\
S|_{V_{3}} &=&\left( 
\begin{array}{ccc}
\frac{(1-z)(1+zq)}{(sz+q)(sz+q^{-1})} & \frac{q(w-1)(sq+1)}{(sz+q)(sz+q^{-1})%
} & \frac{(s+q)(s+q^{-1})}{(sz+q)(sz+q^{-1})} \\ 
\frac{(t+q)(1-z)z}{(sz+q)(sz+q^{-1})} & \frac{w(1+s)(1+t)-tq^{-1}-w^{2}sq}{%
(sz+q)(sz+q^{-1})} & \frac{q(s+q)(sz-t)}{(sz+q)(sz+q^{-1})} \\ 
\frac{(t+q)(t+q^{-1})z^{2}}{(sz+q)(sz+q^{-1})} & \frac{(t+q^{-1})(sz-t)z}{%
(sz+q)(sz+q^{-1})} & \frac{(sz-t)(sz+tq^{-1})}{(sz+q)(sz+q^{-1})}%
\end{array}%
\right) \;.
\end{eqnarray*}

\end{document}